# Multimedia Applications of Multiprocessor Systems-on-Chips

Wayne Wolf
*Department of Electrical Engineering*
*Princeton University*

**Abstract**

This paper surveys the characteristics of multimedia systems. Multimedia applications today are dominated by compression and decompression, but multimedia devices must also implement many other functions such as security and file management. We introduce some basic concepts of multimedia algorithms and the larger set of functions that multimedia systems-on-chips must implement.

## 1 Introduction

Multimedia systems constitue a huge application space for systems-on-chips. Multimedia underlies many common devices for entertainment and and business applications. Because these markets are so large, they require system-on-chip implementations to be successful.

Multimedia algorithms are surprisingly complex. A great deal of effort has gone into algorithms that efficiently compress data, provide high image/sound quality, *etc*. Multimedia standards deploy combinations of many types of algorithms to obtain high-quality results.

Multimedia systems are also complex beyond the needs of the algorithms themselves. A multimedia device must perform many other functions—communications, file management, security, *etc*.—in order to perform all its duties. These added duties must also be supported by systems-on-chips for multimedia.

In the next section, we will briefly survey the types of applications that may see implementation in multiprocessor systems-on-chips. We will then look at video and audio compression, the dominant multimedia applications today. We will briefly consider content analysis, an emerging application. We will then look at digital rights management and other functions that must be performed by multimedia systems.

## 2 Multimedia Applications for Systems-on-Chips

Systems-on-chips designed specifically for multimedia are targeted at consumer applications, where cost and power are critical. Most of today's consumer multimedia applications are based on compression. Even within that relatively constrained algorithmic domain, consumer multimedia devices cover a broad range of cost/performance/power points:

- multimedia-enabled cell phones;
- digital audio players;
- digital set-top boxes;
- digital video recorders;
- digital video cameras.

Digital video recorders are an example of a multimedia analysis system—they may, for example, skip commercials or skip to the next scene based upon an analysis of the content. We should expect to see a wider range of analysis-oriented products as more computational power becomes available and consumers feel the need to use tools to manage their content.

Video compression algorithms symmetric and asymmetric applications. A symmetric compression system is designed to require roughly equal computational power from both the sender and receiver. Videoconferencing is a classic example of this scenario, in which each terminal must both transmit and receive. Asymmetric systems put more effort into encoding to simplify the decoder. Broadcast systems, in which a complex transmitter supplies content to many simpler receivers, is an example of an asymmetric system.

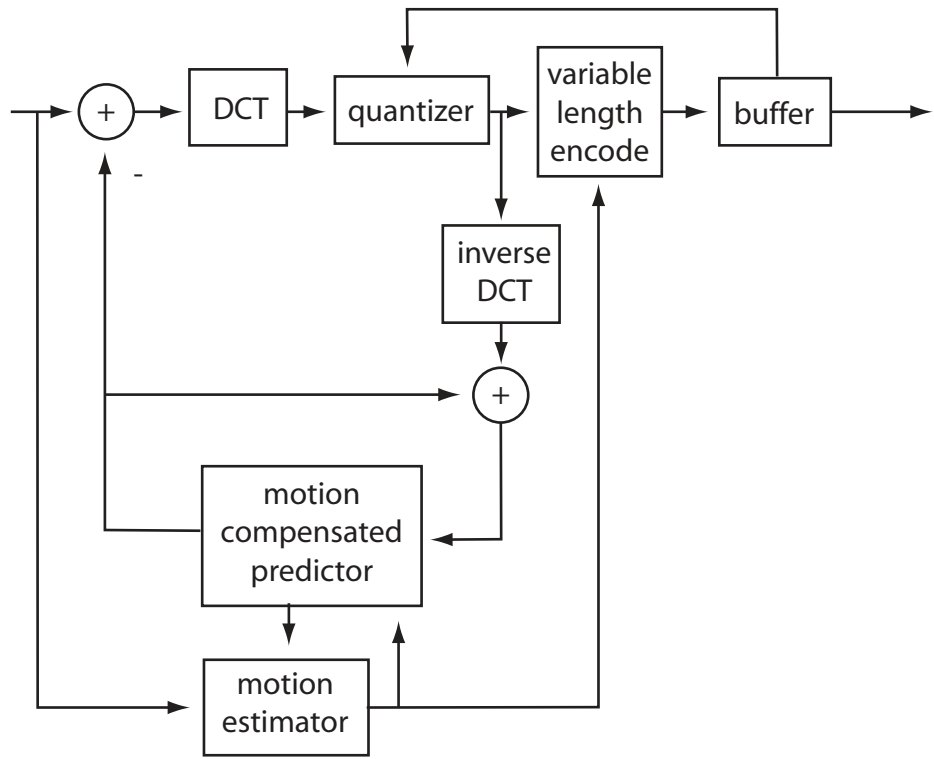

FIGURE 1. Structure of a video encoder.

## 3 Video Compression

Video compression is lossy encoding---information is removed and the reconstructed signal is not identical to the original signal. The structure of a video encoder is shown in Figure 1. Modern video compression systems rely on three major techniques [Has97]:

- The discrete cosine transform (DCT) is used to select details to remove.
- Lossless encoding, particularly Huffman-style encoding, is used to remove entropy from the final data stream sent to the decoder.
- Motion estimation and compensation are used to determine the representation between frames

The first two are also used in image compression, while the third is unique to video.

The discrete cosine transform helps the encoder identify information in the frame to be thrown away. It is a frequency transform with the advantage that a 2-D DCT can be computed from two 1-D DCTs. The DCT itself does not fundamentally reduce the amount of information, but it does separate that information into spatial frequencies. The higher spatial frequencies represent finer detail that is eliminated first.

Motion estimation and compensation allow one frame to be described relative to another. Motion estimation compares part of one frame to a reference frame and determines what motion would cause the selected part to appear in the reference frame. Motion compensation at the receiver then applies that motion vector to reconstruct the frame. The receiver must hold the reference frame in its memory but motion estimation/compensation greatly reduce the number of bits required to represent the video sequence.

Wavelets are a frequency representation that has come into common use in the past several years. Wavelets represent the frequency content hierarchically and do not suffer from the edge artifacts common to DCT-based encoding. Wavelets been incorporated into JPEG2000 for image encoding.

One problem with the multiplicity of standards is transcoding. Since different devices may use different compression standards, content must be recoded to be used on a different device. Because encoding is lossy, each generation of transcoding reduces image quality.

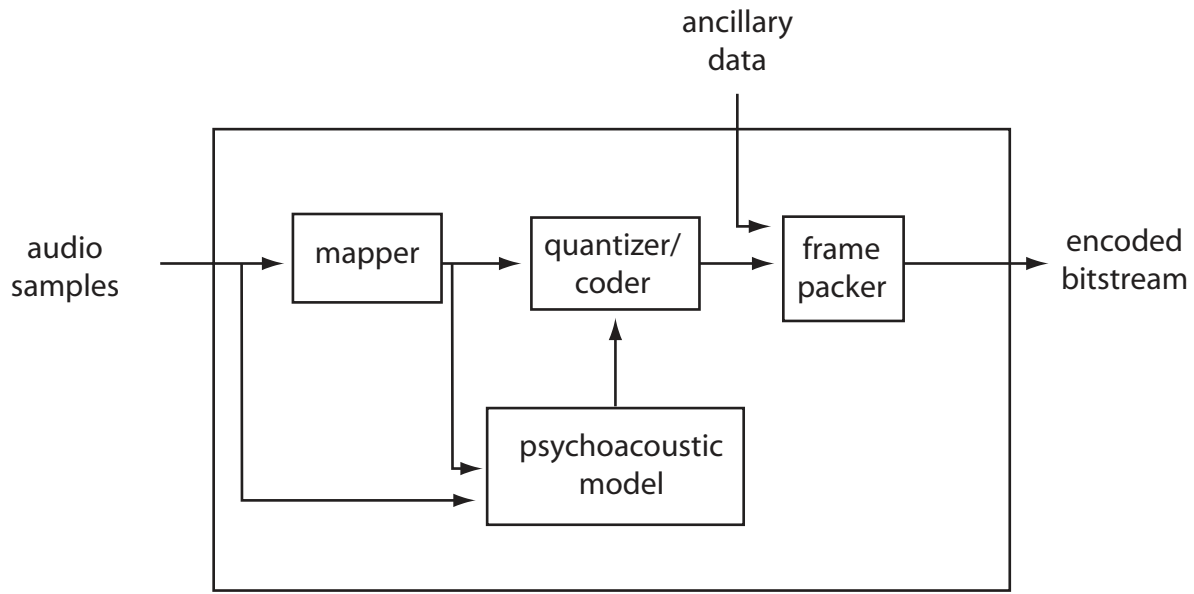

FIGURE 2. The structure of an MPEG-1 audio encoder.

## 4 Audio Compression

The GSM cellular telephony standard uses an audio compression method called Regular Pulse Excitation-Long Term Predictor (RPE-LTP). This method uses a fairly simple model of the voice to encode speech. Speech is often divided into two types of sounds: voiced, which is periodic; and unvoiced, which has broader frequency content. These two types of sound can be generated filtering a combination of glottal resonance and noise. The RPE-LTP encoder generates filter coefficients that can be used at the receiver to generate the required sound from the combination of these two sources.

The well-known audio standard MP3 is shorthand for MPEG-1 Layer 3; the MPEG-1 standard [ISO93] included three layers of audio compression, of which Layer 3 is the most sophisticated. MPEG audio compression methods are best thought of in terms of hearing rather than sound generation as with RPE-LTP.

The structure of an MPEG-1 audio encoder is shown in Figure 2. MP3 uses a combination of subband coding and a psychoacoustic model to compress the audio stream. The psychoacoustic model, because it is relies on the characteristics of the human auditory system, it is not limited to speech. A key psychoacoustic mechanism exploited by compression is masking—when one tone is heard, followed by another tone at a nearby frequency, the second tone cannot be heard for some interval. Human experimentation has determined what time and frequency combinations will be masked. The encoder can eliminate masked tones to reduce the amount of information that is sent to the decoder.

Other standards for audio compression include Dolby Digital (TM), dtx, (TM), and Ogg Vorbis.

## 5 Content Analysis

Content analysis tools use characteristics of the multimedia material to classify the material either as a whole or into its constituent components. As consumers' collections of multimedia content grow, content analysis is becoming increasingly popular as a way to help manage that content.

Audio content analysis has been used to categorize and search for music. Algorithms have had some success in categorizing music into categories and identifying salient features of the music. That information can then be used to recommend similar pieces of music. This type of analysis is generally conducted off-line on a server.

Simple forms of video content analysis has been implemented in consumer devices. The Replay (TM) digital video recorder, for example, automatically identifies com-mercials and skips them. Replay uses black frames between programs and commercials to identify television. Early VCR add-ons identified commercials using the color burst, under the assumption that many movies on broadcast TV were black-and-white while the commercials were in color. A number of research groups have developed algorithms that can parse various types of television content into segments. Such algorithms would allow a viewer to skip an interview segment, for example, and move into the next part of the program.

## 6  Digital Rights Management

Digital rights management (DRM) encompasses all the operations necessary to enforce copyright and license agreements. Digital rights management uses encryption as a tool but it affects the system architecture from user interface to file management.

The DRM system enforces a set of rights that are defined relative to users and to content. Rights may take a number of forms:

- The ability to play certain titles.
- The number of times that a title may be played.
- The right to play a title on more than one device.
- The time period during which the title may be played.

The DRM system may require access to the Internet to be effective. In other cases, DRM may hold rights markers that can be updated over the Internet but do not require a connection for verification.

The playback device must be able not only to perform the authorization transaction but also to play back the content in such a way that the authorizations are not easily subverted. For example, a playback device may be architected to provide only analog output at the pins to prevent direct copying of unencoded digital content.

## 7  Support

Today's multimedia devices must support a wide variety of functions beyond the multimedia algorithms themselves:

- Multimedia cell phones must clearly provide cell phone functionality.
- Many devices must provide Internet access.
- Set-top boxes must provide program guides, pay-per-view authorization, etc.
- Devices with local storage, such as personal audio players or digital video recorders, must provide file systems.

The applications of Internet access vary across devices. Some use the Internet for limited purposes, such as content access or DRM. These devices can make use of the small IP stacks that have been developed over the past several years. Other devices are intended to operate as network devices and to support a variety of transactions across the network. Network-oriented consumer devices often use Java in the upper layers to ease development and allow code to be transported across the network.

Devices with local libraries must have their own file systems. Portable MP3 players and digital video recorders are examples of devices with internal file systems. MP3-enabled CD players are a particularly interesting case since the files are created outside the player. A CD/MP3 player must be able to handle a wide variety of directory structures, file names, etc. A device that creates its own files has somewhat more control over the characteristics of those files. But these file systems must still incorporate the major characteristics of modern file systems: large file sizes, non-sequential allocation of blocks, etc.

Unlike magnetic disk drives, who bundle their control with the drive, DVD recorders and players must control their drives using complex digital filters. The control requires real-time processing at high rates and the control laws are generally adapted to the particular mechanism being used.

## 8  Conclusions

Multimedia applications, such as audio and video compression, are sophisticated collections multiple algorithms. Furthermore, these applications are part of larger systems that must incorporate many features of general-purpose computing systems but at low cost and low power. We should expect both the real-time and background computations required of multimedia systems to become even more complex over the next decade.

## References

[Has97] Barry G. Haskell, Atul Puri, and Arun N. Netravali, *Digital Video: An Introduction to MPEG-2*, Chapman and Hall, 1997.

[ISO93] ISO/IEC, International Standard ISO/IEC 11172-3, *Information Technology—Coding of Moving Pictures and Associated Audio for Digital Storage Media at up to about 1.5 MBit/sec, Part 3: Audio*, 1993.